\documentclass[twocolumn,showpacs,preprintnumbers,amsmath,amssymb]{revtex4}

\usepackage{graphicx}
\usepackage{dcolumn}

\usepackage{mathptmx, courier, pifont}
\usepackage[scaled=0.92]{helvet}
\usepackage[T1]{fontenc}
\usepackage{textcomp}

\usepackage{bm}

\newcommand{\singlefig}[6]{%
\begin{figure} \vspace{#3}%
\includegraphics*[scale=#5]{#2}%
\caption{\label{fig:#1} #6}%
\vspace{#4}%
\end{figure}}

\newcommand{\tsub}[1]{_{\mbox{\scriptsize#1}}}

\begin{document}


\title
{ Generalizing the Cooper-Pair Instability to Doped Mott Insulators
}

\author
{Mike Guidry$^{(1)}$, Yang Sun$^{(2)}$, and Cheng-Li Wu$^{(3)}$ }
\affiliation{ $^{(1)}$Department of Physics and Astronomy,
University of Tennessee, Knoxville, Tennessee 37996--1200
\\ $^{(2)}$Department of Physics, Shanghai Jiao Tong University,
Shanghai 200240, People's Republic of China
\\ $^{(3)}$Physics Department, Chung Yuan Christian University,
Chung-Li, Taiwan 320, ROC
}

\date{\today}

\begin{abstract}

Copper oxides become superconductors rapidly upon doping with
electron holes, suggesting a fundamental pairing instability. The
Cooper mechanism explains normal superconductivity as an instability
of a fermi-liquid state, but high-temperature superconductors derive
from a Mott-insulator normal state, not a fermi liquid. We show that
precocity to pair condensation with doping is a natural property of
competing antiferromagnetism and $d$-wave superconductivity on a
singly-occupied lattice, thus generalizing the Cooper instability to
doped Mott insulators, with significant implications for the
high-temperature superconducting mechanism.

\end{abstract}


\maketitle

Understanding cuprate high-temperature superconductors is
complicated by unusual properties of the normal state and how  this
state becomes superconducting with doping  \cite{gapOnset}.  Band
theory suggests that cuprates at half lattice filling should be
metals, but they are instead insulators with antiferromagnetic (AF)
properties.  This behavior is thought to result from a
Mott-insulator normal state, where the insulator properties follow
from strong on-site Coulomb repulsion rather than band-filling
properties. Upon doping the normal states with electron holes, there
is a rapid transition to a superconducting (SC) state, with evidence
for a pairing gap at zero temperature typically appearing for about
3--5\% hole density per copper site in the copper--oxygen plane.  In
addition, there is strong evidence at low to intermediate doping for
a partial energy gap at temperatures above the SC transition
temperature $T\tsub c$ that is termed a pseudogap (PG), with the
size of the SC gap and PG having opposite doping dependence at low
doping \cite{pseudogap}.

Parent states of normal superconductors are fermi liquids
(strongly-interacting systems having excitations in one-to-one
correspondence with the excitations of a non-interacting fermi gas).
Normal superconductors are described by Bardeen--Cooper--Schrieffer
(BCS) theory \cite{BCS57}, and result from condensation of
zero-spin, zero-momentum fermion pairs into a new collective state
with long-range coherence of the wavefunction. The key to
understanding normal superconductivity was the demonstration by
Cooper \cite{coop56} that normal fermi liquids possess a fundamental
instability:  an electron pair above a filled fermi sea can form a
bound state for {\em vanishingly small attractive interaction.}  In
normal superconductors the attraction is provided by interactions
with lattice phonons, which bind weakly over a limited frequency
range because electrons and the lattice have different response
times. However, it is the Cooper instability, not the microscopic
origin of the attractive interaction, that is most fundamental: a
weak electron--electron interaction alone cannot produce a
superconducting state, but the Cooper instability can (in principle)
produce a superconducting state for {\em any} weakly-attractive
interaction.

The rapid onset of superconductivity in high-$T\tsub c$ compounds
with hole doping suggests a fundamental instability against pair
condensation, but it is difficult to understand this (and the
appearance of PG states) within the standard BCS framework because
the superconductor appears to derive from a Mott insulator, not a
normal fermi liquid.  Just as for normal superconductivity, we
believe that the key to understanding high-temperature
superconductivity is not the attractive interaction leading to pair
binding (as important as that is), but rather the nature of the
instability that produces the superconducting state.  Since at
larger doping the high-temperature superconducting state exhibits
many properties of a normal BCS superconductor (but with $d$-wave
pairs), this instability must reduce to the Cooper instability at
larger doping, but evolve into something more complex at lower
doping where the normal state approaches a Mott insulator and a PG
exists above the SC transition temperature.

To account for rapid onset of superconductivity with hole-doping,
Laughlin \cite{laug02} (see also \cite{zhan03,cole03}) proposed a
modified Hamiltonian with an  attractive term  that partially
overcomes the on-site repulsion.  Then the insulator at half filling
is actually a ``thin, ghostly superconductor'', which fails to
superconduct only because its long-range order is disrupted at very
low doping, ostensibly by fluctuations due to low superfluid
density.   This proposed new  state is termed a {\em gossamer
superconductor.} This idea  might provide a justification for the
resonating valence bond (RVB) state \cite{RVB}, which assumes
implicitly  that quantum antiferromagnets should exhibit
superconductivity, even though cuprate ground states at half-filling
appear to be best described as an insulating state with long-range
AF order and no superconductivity
\cite{vakn87,tran88,chak88,rege88,dago94}.  In the RVB model it is
usually assumed that the long-range N\'{e}el order of the ground
state at exactly half filling is replaced quickly by the RVB
spin-liquid ground state upon hole-doping, with details lacking.
Laughlin \cite{laug02} contends that the real issue for validity of
the RVB picture is not whether all quantum antiferromagnets are
secretly superconductors, but whether some  are.  The gossamer
superconductor is then proposed as a second kind of
antiferromagnetism---distinguished by a small background superfluid
density---that is the true normal state in the cuprates, and is the
harbinger of a spin-liquid RVB ground state for low hole-doping.

The gossamer state has desirable properties but is created by hand:
a strong attractive term is added to the Hamiltonian, which
justifies modifying Gutzwiller projectors such that they only
partially suppress double occupancy \cite{zhan03}. We shall show
that a model implementing competition of $d$-wave pairing with
antiferromagnetic correlations on a lattice with no double occupancy
has Mott insulator properties at half filling, but is unstable
toward developing a finite singlet pairing gap under infinitesimal
hole-doping.  Thus, we shall argue that many features motivating the
idea of gossamer superconductivity are natural consequences of AF
and SC competition on a lattice having strict no double occupancy at
half filling.  We shall argue further that a pseudogap with correct
properties is a natural consequence of the same theory, thus
accounting for both the precocious onset of a pairing gap and the
appearance of pseudogaps at low doping in the cuprates. Finally, we
shall discuss the implications of these results for gossamer
superconductivity and for the RVB model.

We wish to solve for the doping and temperature dependence of
observables in a theory that incorporates on an equal footing
$d$-wave superconductivity and antiferromagnetism. To do so, we
shall employ the tools of Lie algebras, Lie groups, and generalized
coherent states \cite{guid99,lawu03,guid04,sun05,gui07b,zhan90}.  To
construct a Hamiltonian embodying these degrees of freedom and
expected conservation laws for charge and spin in the many-body
wavefunction, we require at a minimum three staggered magnetization
operators $\vec {\cal Q}$ to describe AF, creation and annihilation
operators $D^\dagger$ and $D$ for $d$-wave singlet pairs and a
charge operator $M$ to describe superconductivity, and three spin
operators $\vec S$ to impose spin conservation.

However, this set of 9 operators is physically incomplete since
scattering of singlet pairs (antiparallel spins on adjacent sites)
from the AF particle--hole degrees of freedom can produce triplet
pairs (parallel spins on adjacent sites), which are not part of the
operator set.  The mathematical statement of this incompleteness is
that the operator set $\{\vec {\cal Q}, D^\dagger, D, M, \vec S\}$
does not close a Lie algebra under commutation. As demonstrated in
Refs.\ \cite{guid99,lawu03,sun05}, a (minimally) complete operator
set results if we add to these operators the six triplet pair
operators $\vec \pi^\dagger$ and $\vec \pi$.  Then the set of 15
operators $\{\vec {\cal Q}, D^\dagger, D, \vec \pi^\dagger, \vec
\pi, M, \vec S\}$ closes the Lie algebra SU(4).  The explicit forms
for these operators in both momentum and coordinate space, and the
corresponding SU(4) commutation algebra, may be found in Refs.\
\cite{guid99,lawu03,sun05}.

A critical feature of this symmetry structure is that the SU(4)
algebra closes only if the 2-dimensional lattice on which the
generators are defined has no doubly-occupied sites \cite{guid04}.
Thus, the SU(4) algebra embodies the minimal theory that describes
AF and $d$-wave SC competition through a many-body wavefunction that
conserves charge and spin, and that has no components corresponding
to double site occupancy on the lattice.  The Hamiltonian restricted
to one-body and two-body terms is unique, with the general form
\begin{eqnarray}
H = H_0 -G_0 D^\dag D
-G_1\vec{\pi}^\dag\cdot\vec{\pi}
- \chi\vec{\cal Q}_\cdot\vec{\cal Q}+\kappa\vec{S}\cdot\vec{S} ,
\label{Hsu4}
\end{eqnarray}
where $G_0$, $G_1$, $\chi$, and $\kappa$ are effective interaction
strengths and $H_0$ is the single-particle energy. The  $T=0$ ground
state corresponds to a superposition of singlet and triplet fermion
pairs.

\singlefig
{esurfaces}
{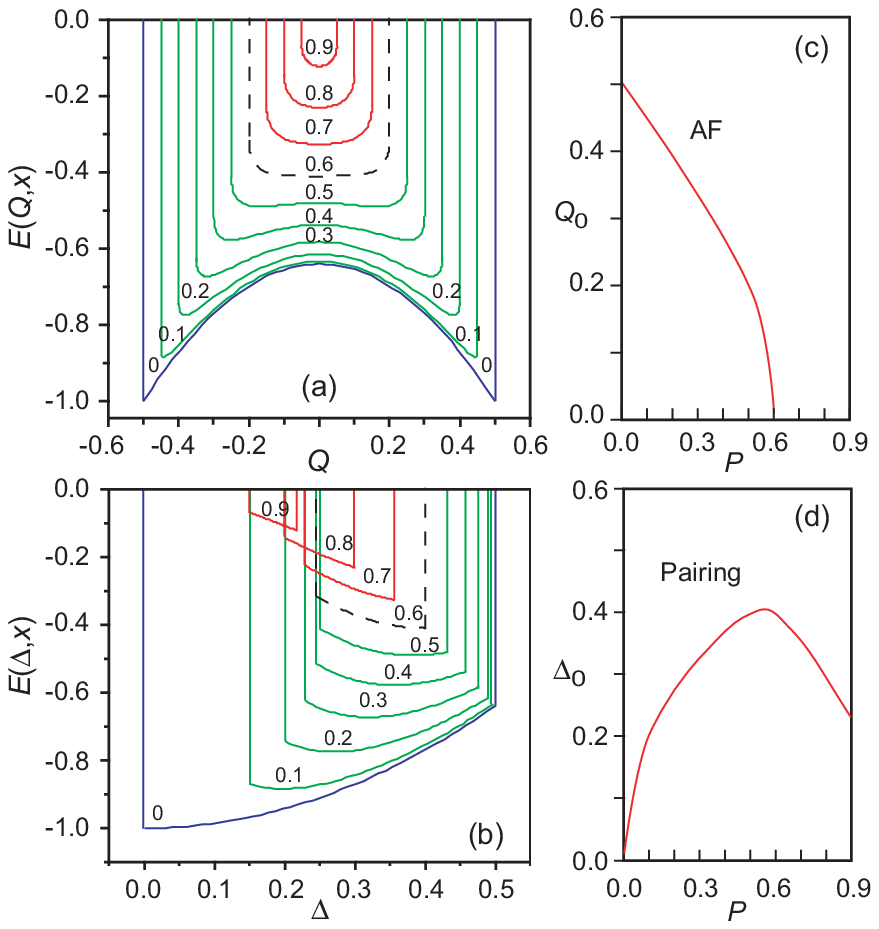}
{0pt} {0pt} {0.94} {Total energy vs.\ (a)~AF correlation $Q$ and
(b)~SC correlation $\Delta$; curves labeled by hole-doping $x\simeq
4P$. Energy in units of $\chi\Omega^2/4$, with $\chi$ the AF
coupling strength. The dashed line indicates the critical doping
$x=x_q$ (see Ref.\ \cite{sun05}); red denotes SC; blue denotes AF;
green denotes AF + SC favoring energy surfaces. Figs.~(c) and (d)
indicate the position of the energy minimum in $Q$ and $\Delta$,
respectively.}

We shall solve for the ground-state properties of this theory using
generalized coherent states \cite{zhan90}.  Our immediate interest
is the ground-state total energy surface, which is the expectation
value of the Hamiltonian in the ground coherent state (with total
spin $S=0$).  The formalism for constructing the SU(4) coherent
state and the ground state energy surface has been developed
extensively in Refs.\ \cite{guid99,lawu03,sun05}, to which we refer
for details.  The overall SU(4) symmetry may be used to eliminate
the $\vec{\pi}^\dag\cdot\vec{\pi}$ term from the Hamiltonian
(\ref{Hsu4}), leaving an energy surface that is a function of order
parameters for singlet pairing and antiferromagnetism, with the
hole-doping as a control parameter.

Figure \ref{fig:esurfaces} illustrates the SU(4) total energy
surface in coherent state approximation as a function of an AF order
parameter $Q = \langle \vec{\cal Q}_\cdot\vec{\cal Q}\rangle^{1/2}/
\Omega$ and an SC order parameter $\Delta = \langle D^\dagger
D\rangle^{1/2}/\Omega$ (where $\Omega$ is the maximum number of
doped holes that can form coherent pairs, assuming the half-filled
lattice as vacuum), as doping $x\simeq 4P$ (for $P$ holes per copper
lattice site) is varied. The explicit expression for the energy
surface is $ E=-\chi \Omega^2[(1-x_q^2)\Delta^2 + Q^2] $, where
\cite{lawu03}
$$
\Delta = \tfrac12 [\tfrac14-(Q-x/2)^2]^{1/2}
+ \tfrac12 [\tfrac14-(Q+x/2)^2]^{1/2},
$$
and $x_q$ is the critical doping at which AF correlations vanish
(see Fig.\ \ref{fig:esurfaces} and Ref.\ \cite{sun05}). Vertical
lines bounding the curves for different doping in Fig.\
\ref{fig:esurfaces} represent the contraints
$$
|Q| \le \tfrac12(1-x) \qquad  [x(1-x)]^{1/2} \le 2\Delta \le
(1-x^2)^{1/2}
$$
that result from SU(4) symmetry within a finite valence space.
Specifically, $|Q|$ must lie between 0 and $n/2$ ($n$ is electron
number) because of the number of spins available,  and SU(4)
symmetry then relates this constraint on $Q$ to the one on $\Delta$.

From Fig.\ \ref{fig:esurfaces}, the energy surface at half filling
($x=P=0$) implies a $T=0$ ground state with  AF order but no pairing
order ($Q_0 \ne 0$ and $\Delta_0 = 0$, where the subscript zero
denotes the value  at the minimum of the energy surface).  The
ground state for $x=0$ may be interpreted as an antiferromagnetic
Mott insulator \cite{guid99,lawu03}.  From Fig.\
\ref{fig:esurfaces}(a), the energy surface retains strong AF
character for small $x$ with $Q_0 \ne 0$, but from Fig.\
\ref{fig:esurfaces}(b) the ground state differs qualitatively from
that at half filling even for infinitesimal hole-doping.
Specifically, for any non-zero attractive pairing strength a finite
singlet $d$-wave pairing gap develops spontaneously for {\em any
non-zero $x$,} and  $\Delta_0$ has already increased to half its
value at optimal doping by the time $P \simeq 0.03$ ($x \simeq
0.12$). The rapid change in the expectation value of the pairing
correlation is illustrated further in Fig.~\ref{fig:esurfaces}(d).
We term this behavior {\em precocious pairing}.

The instability against condensing pairs displayed graphically in
Fig.\ \ref{fig:esurfaces} may also be understood analytically. From
the $T=0$ solution for  $\Delta$ given in Eq.\ (24b) of Ref.\
\cite{sun05}, we find
\begin{equation}
    \left. \frac{\partial\Delta}{\partial x} \right|_{x\rightarrow 0} =
        \left. \frac14 \frac{x_q^{-1} -2x}{[x(x_q^{-1} -x)]^{1/2}}
        \right|_{x\rightarrow 0}
        \rightarrow \infty ,
\label{analyticalDelderiv}
\end{equation}
displaying explicitly the pairing instability at $x=0$.

The picture that emerges is that at half filling the lattice is a
Mott insulator with long-range AF order and no pairing gap, but upon
infinitesimal hole-doping  a finite singlet pairing gap and a ground
state that corresponds to strong competition between AF and SC order
appear. This spontaneous development of a finite singlet pairing gap
for infinitesimal hole-doping has been obtained in the
coherent-state approximation subject to SU(4) symmetry.  Since the
SU(4) algebra closes only if the lattice has no double occupation,
this precocious pairing has occurred without invoking double
occupancy. Doping-dependent effective interactions could delay onset
of SC from $P \sim 0$ to a small doping fraction, as observed, since
the pair gap $G_0\Delta$ vanishes if the singlet pairing strength
$G_0$ vanishes, even for large pair correlation $\Delta$; weak SU(4)
symmetry breaking could also play a role in delaying onset of SC.
However, we propose that the pair-condensation instability of the
SU(4) symmetry limit represents the essential physics governing the
emergence of superconductors from doped Mott insulators.

Mathematically, precocious pairing results from SU(4) invariance,
which requires that $Q^2 + \Delta^2 + \Pi^2 = (1-x^2)/4$, where $\Pi
= \langle \vec\pi^\dagger \vec\pi \rangle^{1/2}/\Omega$ is the
triplet pair correlation \cite{sun05}.  But for a pure AF SU(4)
solution $Q^2=\tfrac14 (1-x)^2$, so even the AF limit has finite
pair correlations $\Delta$ and $\pi$ unless $x= 0$. The {\em
physical origin} of precocious pairing is that a minimal model of
antiferromagnetism, $d$-wave pairing, charge, and spin on a
half-filled fermion lattice is unstable to condensing pairs at
non-zero hole doping under a no double occupancy constraint.

These results have several important implications. We shall discuss
three:  (1)~the interpretation of cuprate data at low hole-doping,
(2)~the implications for gossamer superconductivity, and (3)~the
implications for models of the RVB type.

Cuprate data for low doping suggest that normal compounds at half
filling are AF Mott insulators, that a finite pairing gap develops
by $P\simeq 0.05$, and that a pseudogap develops in the underdoped
region having opposite doping dependence than the singlet pairing
gap.  These results are consistent with a Mott insulator state at
half filling that evolves rapidly into a state with a finite singlet
pairing gap at very low hole-doping.  However, since at low doping
both the singlet pairing and AF correlation energies are
substantial, the AF fluctuations prevent development of strong
superconductivity until near optimal doping, where the
zero-temperature AF correlations are completely suppressed by a
quantum phase transition at $x=x_q$.  We have shown that this same
coherent-state SU(4) theory gives a pseudogap having the observed
doping behavior, and that the pseudogap may be interpreted either as
arising from competing AF and SC degrees of freedom, or from
fluctuations of pairing subject to SU(4) constraints  \cite{sun05}.

The results presented here suggest that an inherent instability
toward condensation of Cooper pairs with hole-doping is a natural
consequence of a minimal model of $d$-wave pairing interacting with
AF correlations on a lattice with no double occupancy.  Thus, the
rapid onset of superconductivity with hole-doping in the cuprates
results from an instability that is Cooper-like (instability against
condensing pairs for non-zero attractive pairing interaction), but
for $d$-wave pairs in an AF Mott insulator. Since the SU(4) coherent
state reduces to a $d$-wave BCS state if AF interactions vanish at
finite hole doping, and to an insulating state with long-range AF
order if pairing and hole doping vanish \cite{sun05}, this
represents a self-consistent generalization of the Cooper
instability to doped Mott insulators. Strong interactions violating
no double occupancy, as in the gossamer hypothesis, are not
precluded but do not seem to be necessary for the pairing
instability.

The SU(4) symmetry-limit solutions are {\em exact} solutions of the
original 2-D lattice problem if the effective interaction is known
(see Section III of Ref.\ \cite{guid99}b).  Since our primary result
depends only on the {\em existence} of an effective interaction with
a finite attractive pairing interaction (which can be checked
empirically), it is exact in the dynamical symmetry limits. The
coherent-state energy surface then represents an approximate
mean-field solution valid for arbitrary doping, but the agreement of
the general coherent-state solution with the exact solutions in the
dynamical symmetry limits for ground-state properties gives
confidence that the coherent state solutions carry the correct
energy-surface properties over the entire physical doping range. In
particular, we reiterate that the instability
(\ref{analyticalDelderiv}) occurs for the  pure AF state ($G_0=0$;
$x_q =1$), which is an {\em exact many-body solution.}

The RVB idea has attractive features but the observed state at half
filling is not a spin liquid.  Motivations of RVB models often gloss
over this difficulty with the assumption that the half-filled state
is  in some sense practically a spin liquid, though it looks like an
AF state.  Our results give independent support for a picture
similar to this, but without RVB assumptions:  the SU(4) ground
state at half-filling has the observables of a respectable
antiferromagnetic Mott insulator, but its wavefunction can
reorganize spontaneously into a superconductor when perturbed by a
vanishingly-small hole doping if there is a non-zero pairing
interaction.  For low hole doping this superconductor is strongly
modified by AF correlations.  Below $T\tsub c$ this gives a $d$-wave
superconducting state weakened by AF correlations; for a range of
temperatures above $T\tsub c$ the pairing gap vanishes but strong AF
correlations in a basis of fermion pairs leads to a pseudogap that
may be interpreted either in terms of preformed pairs or as
competing AF and SC order. Finally, the AF competition weakens with
hole doping until the pure superconductor emerges near optimal
doping.

SU(4) coherent states at low doping presumably share many features
with RVB states.  Triplet pairs are essential for a complete set of
operators in the minimal SU(4) model (for example, no double
occupancy is enforced by the SU(4) algebra, which fails to close
without triplet pairs), and a mixture of singlet and triplet pairs
is essential to describe the AF states at half filling in the
highly-truncated SU(4) fermion basis. But the significance of
triplet relative to singlet pairs decreases rapidly with doping
\cite{sun05} and underdoped SU(4) ground states could have
significant overlap with a singlet spin liquid.  Furthermore,  SU(4)
states at low doping lead naturally to a pseudogap that decreases in
size with increased hole-doping and exhibits fermi arcs, in
quantitative accord with data.

The SU(4) coherent state justifies many features of RVB models, but
it has a richer variational wavefunction than a singlet spin liquid
because it accounts even-handedly for both AF and SC on a lattice
with no double occupancy. Conversely, the SU(4) coherent-state model
is simpler in many respects than RVB models because SC and AF are
accounted for quantitatively in a minimal theory having only
(dressed) electron degrees of freedom:  there are no pair bosons, no
gauge fields, and no spinons or holons (which have formal
justification in one dimension, but are less obviously justified in
higher dimensions, and for which there is little direct evidence in
cuprate superconductors). The SU(4) coherent state represents a
minimal extension of the BCS formalism to incorporate $d$-wave
pairing in the presence of strong AF correlations and large
effective on-site electron repulsion.  It requires no Gutzwiller
projection because the symmetry enforces no double occupancy on the
lattice.  It exhibits a type of spin--charge separation (see Ref.\
\cite{ande97}), but not through topological spinons and holons:  in
the fermion basis, charge is carried both by singlet fermion hole
pairs having a spin of 0 and charge $-2$, and triplet fermion hole
pairs having a spin of 1 and charge $-2$, but spin is carried solely
by the triplet hole pairs.  As we have discussed in References
\cite{guid99,lawu03,guid04,sun05,gui07b}, these features permit a
model that describes many observed features of the cuprates from
half-filling to the overdoped region in a unified manner.

Finally, we comment on the newly-discovered superconductivity in
iron-based compounds \cite{new1}, where a highest $T\tsub c$ of 55 K
\cite{new2} has already been reached. The SC in these materials
seems unconventional, competes with AF \cite{Dai08}, and has  many
other similarities with the cuprates~\cite{Sun08}. The normal states
are (poor) metals, though nearness to a Mott transition can be
debated. It is unlikely that RVB can provide a natural unified
picture of copper and iron based SC.  However, the generalization of
Cooper pairing presented in this paper can accommodate a Mott
insulator normal state at half lattice filling (as appropriate for
cuprates), but is consistent with a metallic normal state for
valence structures characteristic of iron-based SC. Thus a unified
description of cuprate and iron-based superconductors seems possible
within this framework.

In summary, we have examined a minimal theory of cuprate $d$-wave
superconductivity and antiferromagnetism on a lattice with no double
occupancy.  Total energy surfaces imply ground states that are
antiferromagnetic Mott insulators at half filling, but are unstable
against developing singlet  $d$-wave pairing gaps upon hole-doping,
thereby generalizing the Cooper instability to doped Mott
insulators. Many properties motivating gossamer superconductivity
are explained naturally, without invoking the gossamer hypothesis.
We find support for the assumption of resonating valence bond models
that the state with AF order at half filling would really like to be
a spin-singlet liquid.  However,  the SU(4) coherent state is
simpler to implement, yet contains richer physics, than a
spin-singlet liquid, and accounts systematically for many cuprate
properties across the entire physical doping range. Finally, we
suggested that these ideas have the potential to unify descriptions
of cuprate and new iron-based superconductivity.

Yoichi Ando asked insightful questions that partially motivated this
paper, and Elbio Dagotto, Peng-Cheng Dai, Takeshi Egami, and Thomas
Papenbrock provided useful discussions.

\bibliographystyle{unsrt}

\end{document}